\documentstyle[epsf,11pt]{article}
\def\bibitem#1{}

\def\fullref#1{}
\def\arcsec{{\tt "}}
\newif\ifepsfig{\global\epsfigfalse}
\def\figopticdiag{1}
\def\tavphoto{2}
\def\figpcprog{3}

\def\figpupilmasks{5}
\def\figcoroprf{6}
\def\figspec5774{7}
\begin{document}
%
%\centerline{\Large \bf The ``PISCO'' speckle camera at Pic du Midi 
%observatory%%
\centerline{\large \bf THE ``PISCO'' SPECKLE CAMERA AT PIC DU MIDI 
OBSERVATORY%%
\footnote{Based on observations made 
at T\'elescope Bernard Lyot, Pic du Midi, France}}
\bigskip
\centerline{\large \bf J.-L.~Prieur$^*$, L.~Koechlin$^*$,
 C.~Andr\'e$^*$, G.~Gallou$^*$, C.~Lucuix$^*$}
\medskip
\centerline{\large \sl $^*$Observatoire Midi-Pyr\'en\'ees,
14, Avenue E. Belin, F-31400 Toulouse, France.
             }
%Email: prieur at astro.obs-mip.fr
%
\bigskip
\medskip
\noindent
{\large\bf Abstract:}

\medskip
\noindent
We present a new speckle camera designed and built at Observatoire
Midi-Pyr\'en\'ees. This focal instrument has been used for two years with
the 2-meter Bernard Lyot Telescope  of Pic du Midi observatory. It can be
set in various operating modes: full pupil imaging, masked-pupil imaging,
spectroscopy, wave-front sensor and  stellar coronagraphy, hence its name
``PISCO'' (``Pupil Interferometry Speckle COronagraph").
Restored images of double and triple stars have demonstrated its
capabilities in providing close to diffraction limited 
images (0.06\arcsec$\,$ in V). 
PISCO has been fully tested and is now ready to be used 
by the whole astronomical community.
%
%\keywords{speckle interferometry -- instrumentation.}

%
\bigskip
\noindent{\large\bf 1. Introduction}

\medskip
\penalty 10000
This speckle camera was designed and built between 1991 and 1994
by the ``Aperture synthesis team'' of Observatoire Midi Pyr\'en\'ees (OMP), 
as a new focal instrument for the 2 meter T\'elescope Bernard Lyot (TBL) 
at Pic du Midi. The aim was 
to take advantage of the good seeing quality  of that site.

The optical design was chosen such that 
a pupil plane would be accessible and 
pupil masks could be put into it, 
to allow for aperture synthesis experiments and
simulate telescope arrays such as
the ESO VLTI (European Southern Observatory  Very Large Telescope in the
Interferometric mode) or others. This speckle camera
thus provides appropriate experimental tools for the investigation of
image restoration techniques with the optical telescope interferometric
networks that are currently being built and operated around the world 
(CHARA, COAST, GI3T, IOTA, NPOI, PTI, SUSI, etc, 
see {\it f.i.} 
{\sl Astronomical Interferometry,} Proceedings of SPIE, Vol. 3350,
Kona, 20-28/03/98)

In the following we present an overview of PISCO and its optical concept
(\S 2).  Then we introduce the various operative modes and illustrate
them with observational results (\S 3). 
The detectors that have been used with PISCO  are presented in \S 4.
In \S 5, we discuss the current performances of the instrument in relation
with  the physical limitations of the speckle techniques and the image
restoration methods. In \S 6 we briefly describe the on-going scientific
programs which use PISCO, and conclude about the use of a speckle camera
in the current context of new high angular resolution techniques.

\bigskip
\noindent{\large\bf 2. The PISCO speckle camera}  

\penalty 10000
\medskip
Compared to other speckle cameras (Blazit {\it et al.}, 1977, 
Breckinridge {\it et al.} 1979, Strittmater, 1980, 
Beckers {\it et al.}, 1983, % Steward Obs. differential speckle cam.
Foy, 1988a) our instrument presents the advantage of
a versatility and a full remote control of all its operating modes. 
PISCO offers thus a wide range of possibilities
with fast switching between modes (less than one minute), allowing 
an optimal use of the seeing conditions.

\bigskip
\noindent{\large\sl 2.1. Description}

\penalty 100000
\medskip
The general layout is shown in
Fig.~\figopticdiag.
The external mechanical structure is a rectangular box of
100$\times$40$\times$36~cm$^3$ which is mounted at the
Cassegrain focus of the telescope (Fig.~\tavphoto).

The input image plane (I1) of the telescope is located
200~mm downstream from the front flange of PISCO
(this value is easily adaptable for the 3.6~m CFH or ESO telescopes).
The converging input beam is transformed into a parallel beam by the 
collimating lens (L2), then focused into the image plane (I2)
by the lens (L3). 

Lens (L4) magnifies this image and projects it to the detector
faceplate. The focal length of (L4) can be selected with the wheel GR
that bears  a series of eyepieces and microscope objectives. A 
magnification of at least 20~mas/pixel is needed to obtain a good sampling
for speckle observations at the TBL 
while a lower magnification is used for field acquisition.

%%%%%%%%%%%%%%%%%%%%%%%%%%%%%%%%%%%%%%%%%%%%%%%%%%%%%%%%%%%
%%%%% Fig 1: Optical diagram
\centerline{\epsfysize=8cm\epsfbox{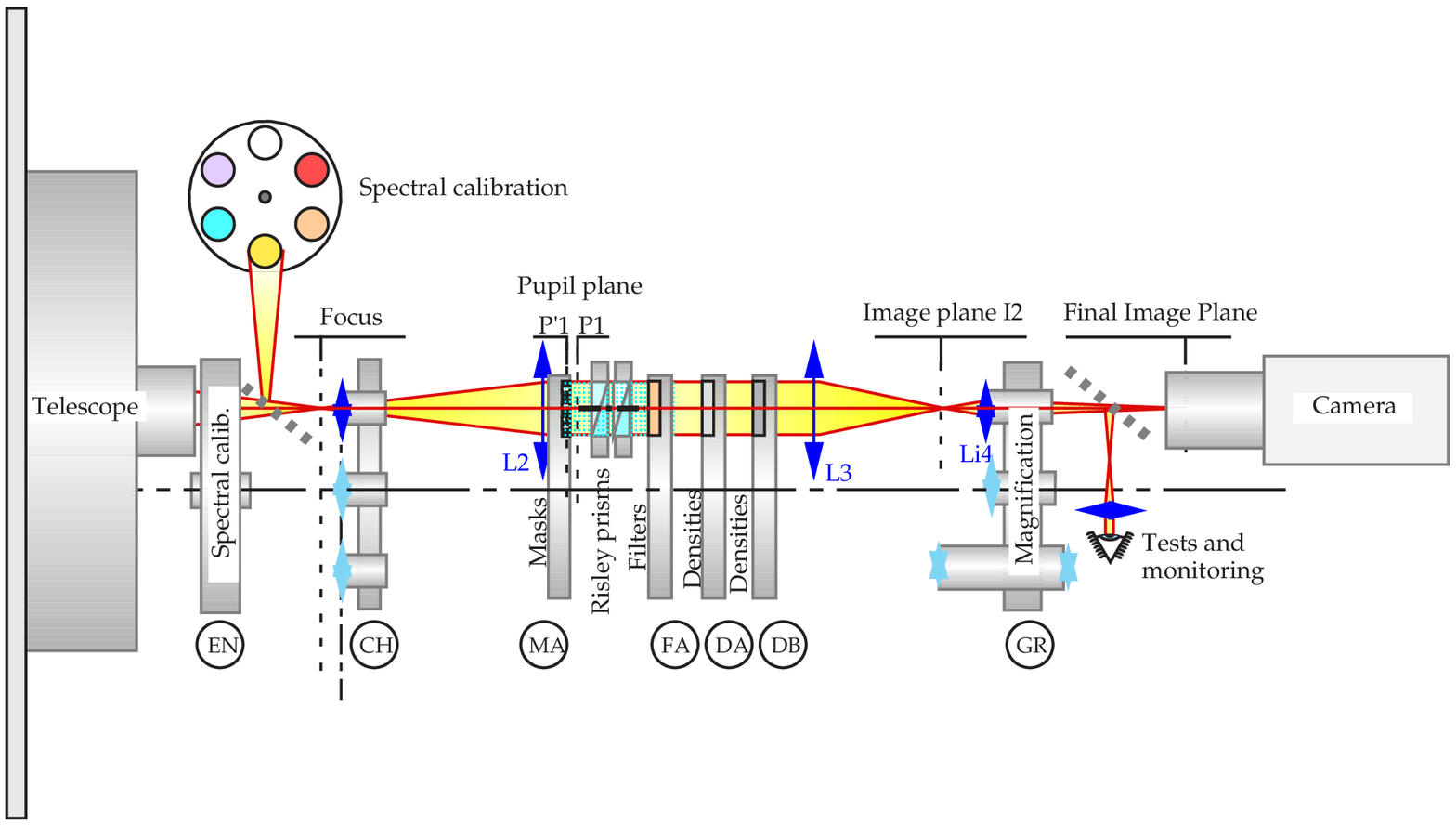}}
\smallskip
\centerline{Figure 1: Optical diagram of PISCO.} 
\bigskip
%%%%%%%%%%%%%%%%%%%%%%%%%%%%%%%%%%%%%%%%%%%%%%%%%%%%%%%%%%%
 
Filters (wheels FA and FB) allow selecting the desired wavelength range
while neutral densities ( wheels DA and DB) are used to adjust 
the light level to the 
(generally poor) dynamic range of the photon-counting detectors 
(see~\S 4).
As will be explained in \S~2.2, 
a set of Risley prisms correct for the atmospheric chromatic dispersion.

%%%%%%%%%%%%%%%%%%%%%%%%%%%%%%%%%%%%%%%%%%%%%%%%%%%%%%%%%%%
%%%%% Fig 2: Photo of PISCO
\centerline{\epsfysize=8cm\epsfbox{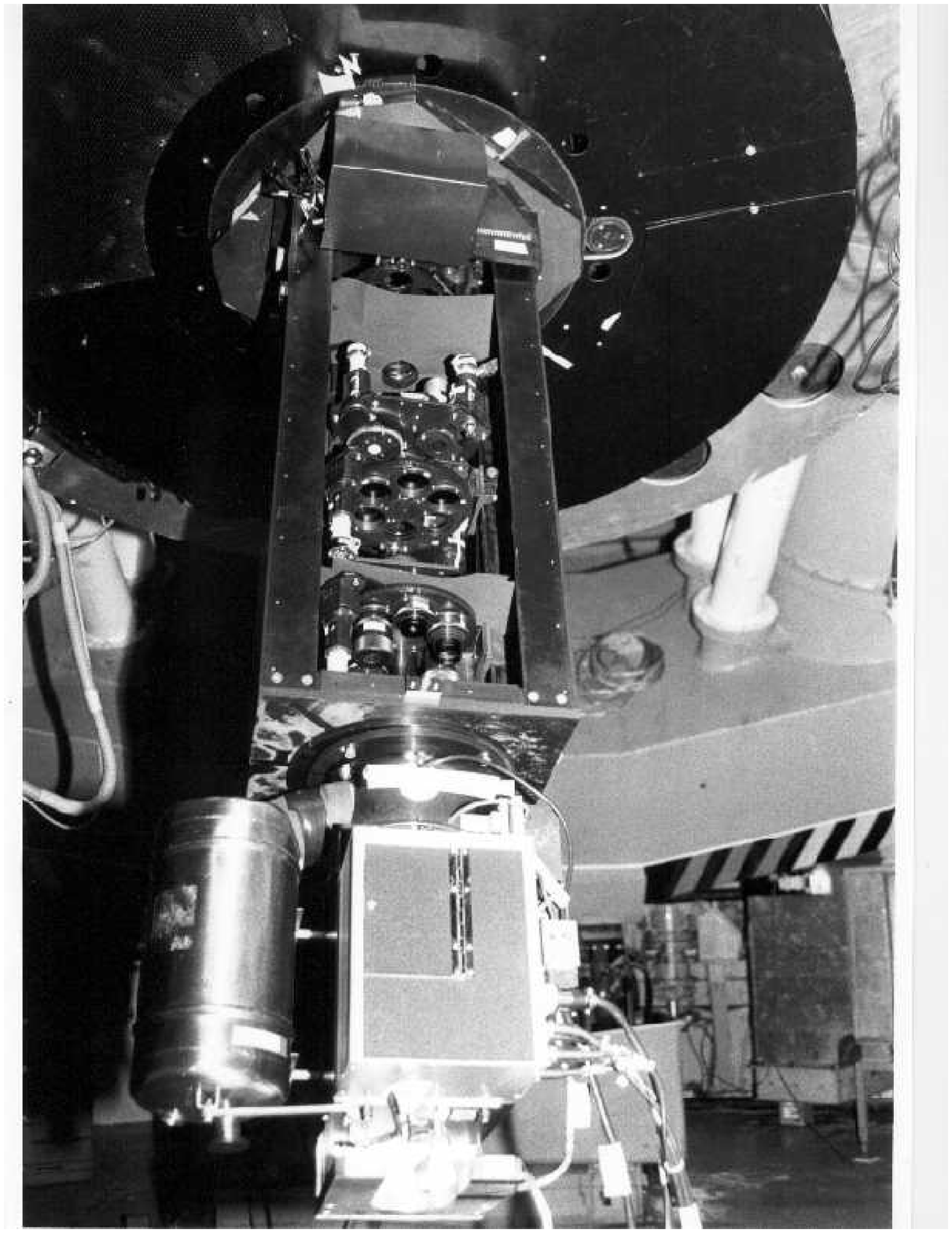}}
\smallskip
\centerline{Figure 2: PISCO and photo-counting detector CP40 at the TBL.}
\bigskip
%%%%%%%%%%%%%%%%%%%%%%%%%%%%%%%%%%%%%%%%%%%%%%%%%%%%%%%%%%%

When a field lens is selected (wheel CH), the pupil plane is 
located in the plane of the wheel MA, where pupil masks 
are available for coronagraphy~(\S 3.3) or 
multi-aperture interferometry~(\S 3.2). 
If low dispersion spectroscopy~(\S 3.4) is wanted,
a grism can be put into the parallel beam (wheel FA).
Wavefront analysis is also possible with a Hartman sensor
by selecting a microlens array in the wheel MA, and 
the pupil imaging mode in wheel GR.

PISCO can then be seen as an optical bench on which 
mounts and wheels can freely move and rotate.
This concept allows a great flexibility 
for future instrumental developments. 
All instrumental functions, 
including wheel positioning and control of the Risley prisms
%(atmosperic dispersion corrector), 
are monitored by a microprocessor and remotely accessible via a RS232 link.

One of us (J.-L. Prieur) has developed a program in the PC/Window environment
to facilitate the remote control of PISCO. All the basic functions are
available (Fig.~\figpcprog) with mouse-driven menus. The program
controls in real time the atmospheric dispersion correction 
according to the telescope position, the filter and the atmosphere 
parameters (\S 2.2). A log file is also produced at the end
of each night with the PISCO setup parameters of all the exposures 
taken during that night.

%%%%%%%%%%%%%%%%%%%%%%%%%%%%%%%%%%%%%%%%%%%%%%%%%%%%%%%%%%%
%%%%% Fig 3: Control panel of the PC program
\centerline{\epsfysize=90mm\epsfbox{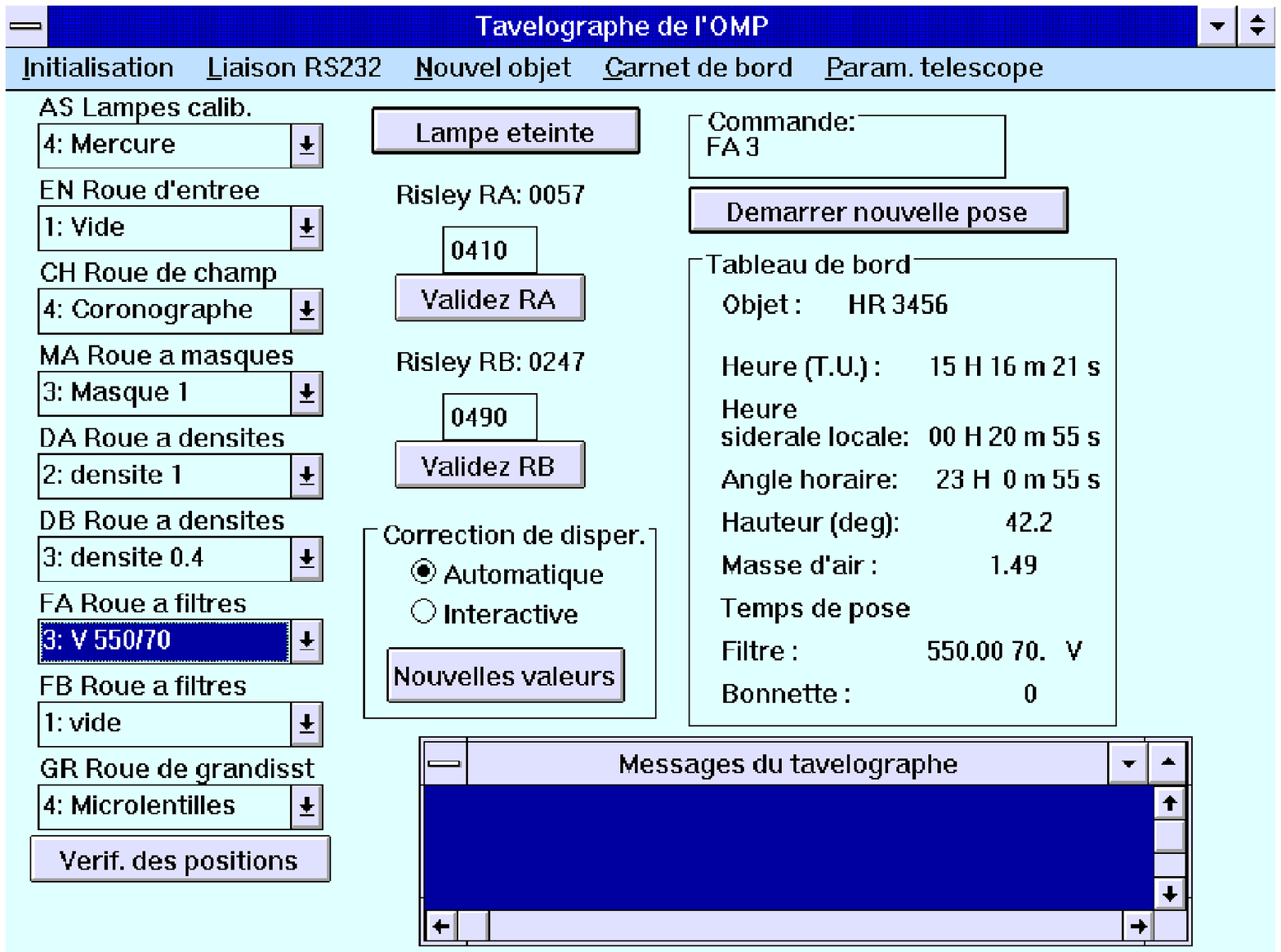}}
\smallskip
{Figure 3: Main panel of the program which 
controls the speckle camera PISCO.}
\bigskip
%%%%%%%%%%%%%%%%%%%%%%%%%%%%%%%%%%%%%%%%%%%%%%%%%%%%%%%%%%%

PISCO was primarily designed to be used  at the Cassegrain focus of TBL,
but it has also been made mechanically and optically compatible with the 
Cassegrain foci of the Canada-France-Hawaii (CFH) and  European Southern
Observatory (ESO) 3.6-m telescopes.

\bigskip
\noindent{\large\sl 2.2. Atmospheric dispersion correction with Risley prisms}

\penalty 10000
\medskip
For an astronomical object observed from the ground
at an elevation different from zenith, the
atmosphere behaves as a dispersive prism 
(see f.i., Simon, 1966).
Polychromatic images are spread into a small vertical spectrum. 
For instance, for a 250~nm bandwidth centered at 500~nm, the typical 
atmospheric dispersion  is 1\arcsec\ for an elevation $h = 60^\circ$ 
and 2\arcsec\  for $h =30^\circ$.

In PISCO, the atmospheric dispersion is
corrected with ``Risley prisms,'' 
which consist in two identical sets of prisms 
(Breckinridge {\it et al.} 1979, Walner, 1990) 
that can be rotated to produce a tunable chromatic dispersion both
in amplitude and direction (Fig.~4a).  

Each set is made of two prisms of different
dispersion law and roof angles, placed in an upside-down
position. These prisms have been designed to have
a null mean deviation, and a dispersion allowing
atmospheric correction from the zenith down to
an elevation of 30$^\circ$ for the 
blue domain which is the most defavorable 
(B filter, centered at 450~nm with a 70~nm bandwidth).
We used the same combination of Shott glasses (F4, SK10) 
as the one used for the Kitt Peak speckle camera 
(Breckinridge {\it et al.}, 1979).
Wallner (1990) found other combinations which
are closer to the atmospheric dispersion
curve  but we chose
the Kitt Peak combination because of its low cost and sufficient efficiency
for our purpose. Our Risley prisms reduce the
residual dispersion down to a level
smaller than 0.01" for every location of the object
in the sky above an elevation of~30$^\circ$, with the
70~nm bandpass B filter which is small compared to the diffraction limit 
of 0.05" in B at the~TBL. 

During the observations, a specially designed program (already
mentionned in \S 2.1)
computes the elevation of the star and the corresponding atmospheric
dispersion using J.C.~Owens' model of atmosphere 
(Owens, 1967, formulae 29--31).
The Risley prisms are then 
dynamically rotated during data acquisition 
to compensate for the atmospheric dispersion.

\bigskip
\noindent{\large\bf 3. The observation modes}

\penalty 10000
\medskip
PISCO can be used in various modes which are selected during the
observations. Switching from one mode to another takes a few mouse
clicks and less than one minute for the motors to set the wheels.

\bigskip
\noindent{\large\sl 3.1. Full pupil speckle imaging}

\penalty 10000
\medskip
In this mode, no pupil masks are used and a high magnification is
selected with wheel GR (Fig.~\figopticdiag). 
It corresponds to the ``conventional'' way of
observing in speckle interferometry. Most speckle cameras
(Breckinridge {\it et al.} 1979, Strittmater, 1980, Foy, 1988a).
offer only this possibility --~or would require many optical changes 
with a full re-calibration of the instrument for other operating modes~--.
By applying bispectral techniques (\S 5.2), we have obtained 
images with an angular resolution close to 
the diffraction limit of the telescope (Fig.~4b). 

%%%%%%%%%%%%%%%%%%%%%%%%%%%%%%%%%%%%%%%%%%%%%%%%%%%%%%%%%%%
\hbox to \hsize{
%%%%% Fig 4a: Atmospheric dispersion correction with Risley prisms
\vbox{\hsize=6cm
\epsfxsize=8cm\epsfbox[0 200 600 700]{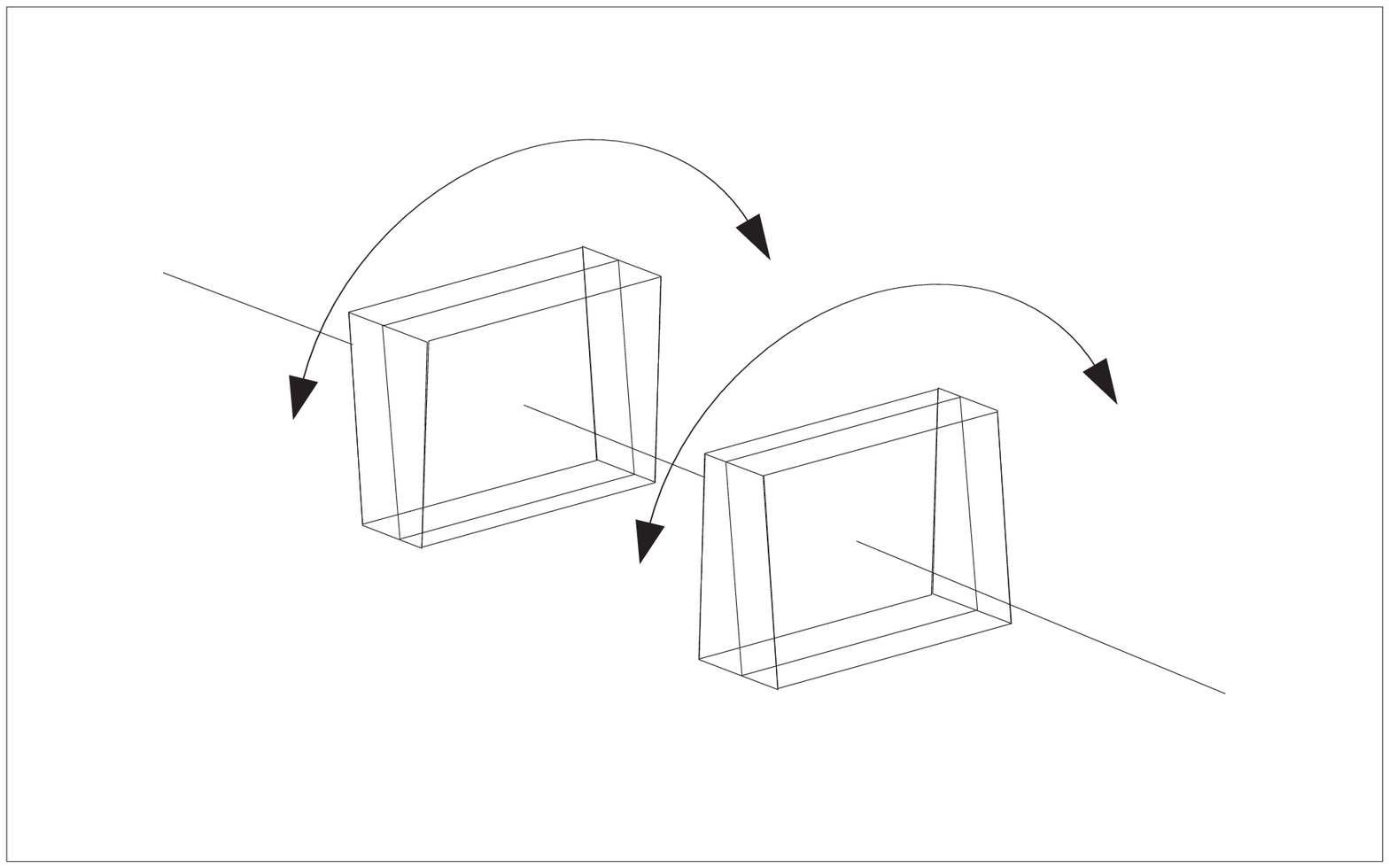}
}
\hfill
%%%%% Fig 4b: Restored image
\vbox{\hsize=6cm
\epsfysize=6cm\epsfbox{2cam_gray.eps}
}
}
\bigskip
\hbox to \hsize{%%%
\parindent=0pt
\hfill
\vtop{\hsize 7cm 
Figure 4a. Risley Prisms to correct for the atmospheric dispersion.}
\hfill
\hskip 1truemm
\hfill
\vtop{\hsize 7cm
Figure 4b. Full pupil speckle imaging:
restored image of the triple star 2~Cam, at the TBL.}
\hfill
}
\bigskip
%%%%%%%%%%%%%%%%%%%%%%%%%%%%%%%%%%%%%%%%%%%%%%%%%%%%%%%%%%%

The full pupil is used and the optical transfer function (OTF)
corresponds to that of the
telescope, and we shall see in the following that another OTF
may be preferred. For instance, the diffraction pattern 
of the telescope spider may pollute the final image and hinder
the detection of faint objects in the vicinity of a bright one.

Another drawback of this OTF is that
low spatial frequencies dominate the transfer function.
In photon counting mode mode, the few available photons 
are then mainly spread in the (less useful) low frequencies.

As the limited dynamic range of detectors such as the CP40 
imposes the use of neutral densities 
to reduce the photon flux for bright objects 
to only a few hundreds per frame(\S 4),
they always work in photon counting mode.
This limits the performances of the restoration methods even for bright
objects.
This is the reason why the next mode could be preferred in that case.

\bigskip
\noindent{\large\sl 3.2. Masked pupil speckle imaging and aperture synthesis}

\penalty 10000
\medskip
By inserting masks into the pupil plane (P1)  (cf. Fig.~\figopticdiag), 
the pupil function (and thus the OTF) can be modified as
desired. For instance the spider diffraction pattern can be removed by
placing a Lyot's mask or a four-hole mask which carefully
avoids the shadow of the spider (see \S 3.3), and telescope arrays
can be simulated by placing a mask with appropriately located 
small holes. 

Pupil masks allow to select a sub-sample of spatial frequencies and more
accurately measure the corresponding complex visibilities since they will
be less attenuated (as the overlap of the fringes is lower); hence, a
better use of the maximum number of photons allowed by the detector. The price
to pay is to perform an interpolation in the Fourier space (aperture
synthesis, see methods in \S 5.1) and complementary observations to make
the process more robust in the case of complex objects.

We made 3 pupil masks by drilling 0.7~mm holes into
a 5-cm  metallic disk, according to some of
the complementary non-redundant networks 
from Golay (1971). They are displayed in Fig.~\figpupilmasks\  
with the corresponding $(u,v)$ coverage.
A successful image of HR~8652 was restored using these masks 
and the aperture synthesis method from Lannes (1989, 1991)
and Anterrieu (1992). These method would work with any
other configuration --~there is no need for complementary
networks~--. We chose these masks because the corresponding 
$(u,v)$~coverage~(Fig.~\figpupilmasks) was rather compact
with only small gaps, which makes image restoration more robust.

The main drawback is a lower limiting magnitude
in the case of small holes, and this method can only be used for
objects with $V < 8$ at the TBL.

%%%%%%%%%%%%%%%%%%%%%%%%%%%%%%%%%%%%%%%%%%%%%%%%%%%%%%%%%%%
\hbox to \hsize{\parindent=0pt
\hfil
\vbox{\hsize=7cm 
%%%%% Fig 5a: Pupil masks (Golay)
\vskip 1truecm
\epsfxsize=8cm\epsfbox{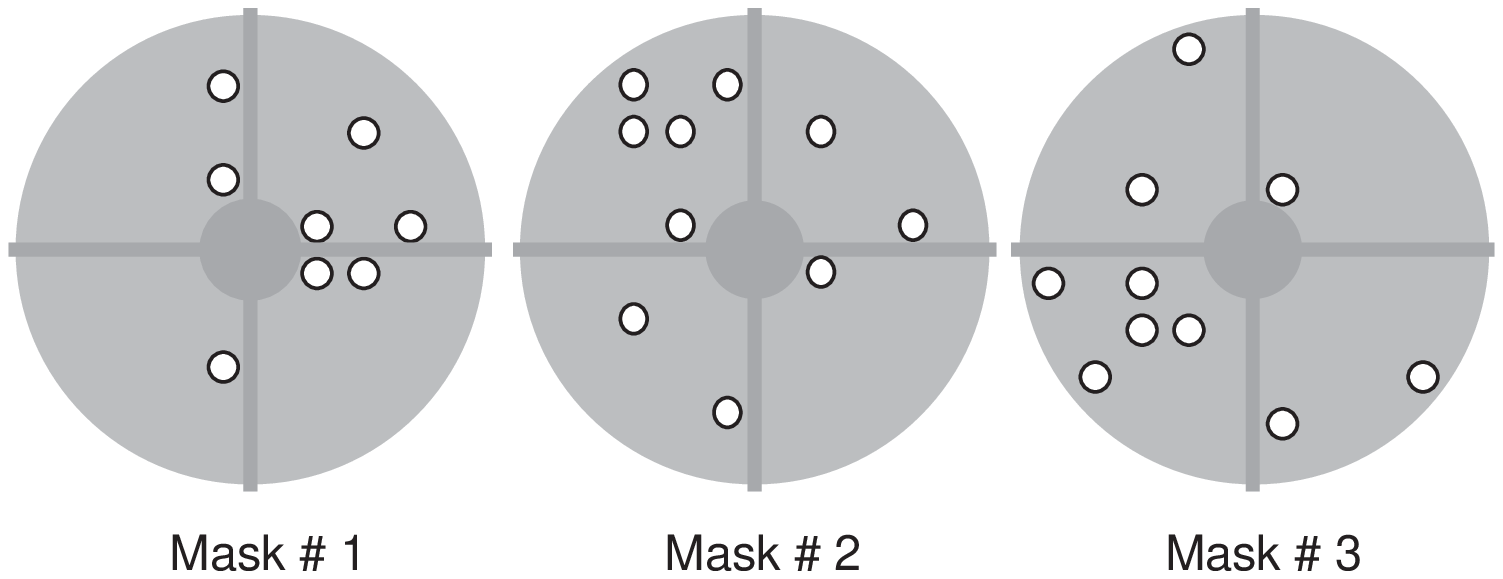}
\vskip 1truecm
}
%%%%%%%%%%%%%%%%%%%%%%%%%%%%%%%%%%%%%%%%%%%%%%%%%%%%%%%%%%%
\hfil
%%%%%%%%%%%%%%%%%%%%%%%%%%%%%%%%%%%%%%%%%%%%%%%%%%%%%%%%%%%
\vbox{\hsize=7cm
%%%%% Fig 5b: (u,v) coverage with pupil masks 
\epsfxsize=5cm\epsfbox[0 0 500 500]{couvuv.eps}
%%%%%%%%%%%%%%%%%%%%%%%%%%%%%%%%%%%%%%%%%%%%%%%%%%%%%%%%%%%
}%%% end of vbox
\hfil
}%%% end of hbox
\smallskip
{Figure 5: Set of pupil masks available in PISCO (left)
and corresponding $(u,v)$ coverage (right).}

\bigskip
\noindent{\large\sl 3.3. Coronagraphic mode}

\penalty 10000
\medskip
PISCO can be used as a Lyot's coronagraph
by putting adequate masks $m_1$ in the entrance image plane (I1)
and $m_2$ in the pupil plane (P1) (wheels EN and MA 
of Fig.~\figopticdiag). This mode was successfully
tested in 1994 with long integrations on a conventional CCD detector.
Speckle imaging from short-exposured frames 
has little interest in this mode since the obturation of the
mask $m_1$ in (I1) needs to be quite large (a few times the FWHM seeing) 
to hide most of the brightness of the central target. This mode would take 
its real advantage with adaptive optics and a small obturation of $m_1$, to
investigate closer to the target (see f.i., some recent developments
with a phase mask in the image plane, Roddier \& Roddier, 1997).

A four-hole pupil mask 
can also be used to suppress the diffraction image of the spider
of the telescope. This reduces the diffusion
of a bright object and allows the detection
of a possible faint close companion or stellar envelopes, as seen 
in Fig.~\figcoroprf. The stellar profile is more 
concentrated when putting this mask. If we normalize the profiles
with the central value, 
the level of the wings has been reduced by a factor larger 
than~3.

%%%%%%%%%%%%%%%%%%%%%%%%%%%%%%%%%%%%%%%%%%%%%%%%%%%%%%%%%%%
%%%%% Fig 6: Four-hole coronagraphy mask
\hbox to \hsize{\parindent=0pt
\hfil
\vbox{\hsize=7cm
%%%%% Fig 6a: 4-hole mask 
\vskip 1truecm
\epsfysize=3cm\epsfbox{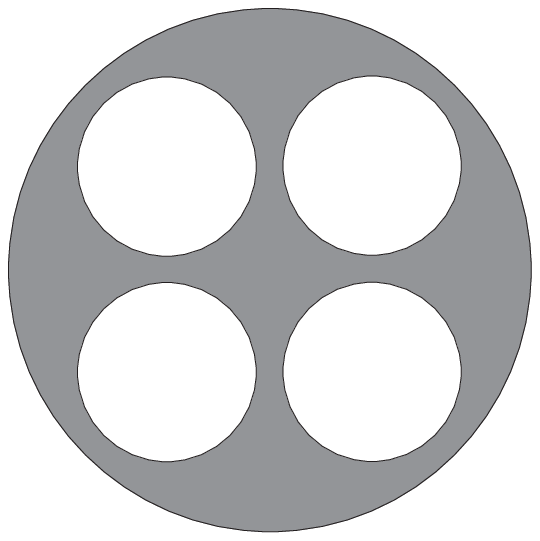}
\vskip 1truecm
}
%%%%%%%%%%%%%%%%%%%%%%%%%%%%%%%%%%%%%%%%%%%%%%%%%%%%%%%%%%%
\hfil
%%%%%%%%%%%%%%%%%%%%%%%%%%%%%%%%%%%%%%%%%%%%%%%%%%%%%%%%%%%
%%%%% Fig 6b: Profile obtained with the coronographic mask
%%% HR 5774 r0031 - 042  r0034 - 0.18
\vbox{\hsize=7cm
\epsfxsize=8.8cm\epsfbox{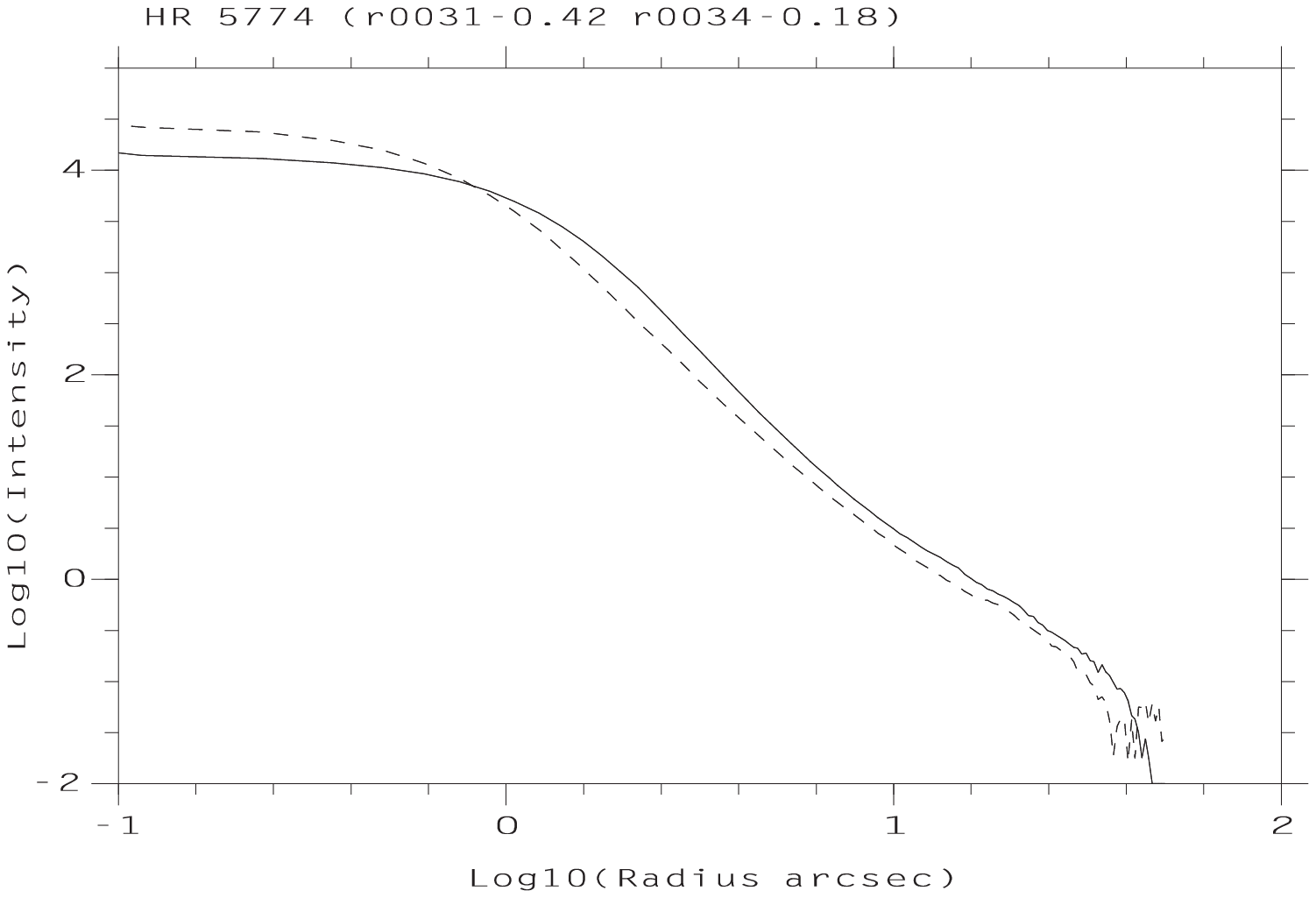}
}%%% end of vbox
\hfil
}%%% end of hbox
\smallskip
{Figure 6: 
Pupil mask used to remove the diffraction pattern
of the spider in the coronagraphy mode (left),
and profile of the star HD5774 (right) with 
(dashed line) and without (solid line)
that mask.} 

%%%%%%%%%%%%%%%%%%%%%%%%%%%%%%%%%%%%%%%%%%%%%%%%%%%%%%%%
%%%%%%%%%%%%%%%%%%%%%%%%%%%%%%%%%%%%%%%%%%%%%%%%%%%%%%%%%%%

\bigskip
\noindent{\large\sl 3.4. High angular resolution spectroscopy}

\penalty 10000
\medskip
Some authors have already shown the feasibility of speckle spectroscopy 
(Weigelt {\it et al.}, 1991, Kuwamura {\it et al.}, 1992) which
has a great interest for the individual study of binary stars or
for determining the physical nature of fine details 
found in speckle imaging. Two possibilities have been used: spectroscopy
with or without a slit in the entrance image plane.

\noindent
-- Both the Hokudai speckle camera at the Okayama 188~cm telescope 
and the Steward Observatory speckle camera with the
spectroscopy module at KPNO used by Kuwamura {\it et al.}, 1992,
worked in {\sl objective prism spectroscopy mode}, i.e., slitless
spectroscopy. In this mode, the resolution is not fixed and changes
as the seeing varies. The spectral calibration is rather difficult 
to perform since it depends upon the position of the object in
the field.
But the main advantage is that all the incoming light is used, 
without any loss. 

\noindent
-- Weigelt {\it et al.}, 1991, proposed a slit spectroscopy setup
which allows a high spectral resolution
and a fixed spectral calibration.
This is the option we chose because we wanted to be able 
to do stellar classification of the
components of binary stars and work with a good spectral calibration.
The main drawback is a loss of sensitivity due to a rejection
of the light by the entrance slit.
%that could be partly solved by an image slicer, which is not
%available yet.

PISCO can be easily converted to a spectrograph
by selecting the grism in the wheel FA and a slit in the
wheel EN in the entrance image plane (I1) (Fig.~\figopticdiag).
It then provides a low dispersion spectrographic mode 
with a spectral range of 350--500~nm, 
and a spectral resolution of $\sim$300 with a
slit of 0.7\arcsec. 
This range was chosen to allow stellar classification
of close binaries with the hydrogen Balmer series. Unfortunately,
the atmospheric turbulence is stronger in the blue domain,
which make images more difficult to restore.
The wavelength calibration can be performed with
calibration spectral lamps in wheel AS (Halogen, Argon, Neon 
and Xenon lamps).

Quick switching between imaging and spectroscopy
is possible since this mode can be remotely selected 
by rotating the wheels. The observing procedure is
the following: 

-- obtain the autocorrelation of the binary star in
the full pupil mode (cf., \S 3.1) and measure the rotation angle 
to align the slit on the direction of the two components.

-- rotate the telescope flange supporting PISCO
and center the object on the slit.

-- switch to the spectroscopic mode and record the data. 

Restoration of high angular resolution information 
in the direction of the entrance slit 
is then done by applying one-dimensional speckle imaging techniques 
to each monochromatic image of the slit (see \S 5.2).

The first spectroscopic observations were
made in 1995. Unfortunately
the poor seeing conditions and the low dynamic range
of the detector did not allow us to restore high resolution
images. We simply obtained long integration spectra (Fig~\figspec5774)
and calibrated the whole instrument with known stars.

%%%%%%%%%%%%%%%%%%%%%%%%%%%%%%%%%%%%%%%%%%%%%%%%%%%%%%%%%%%
%%%%% Fig 7: First spectrum of PISCO
\bigskip
\centerline{\epsfxsize=8.8cm\epsfbox{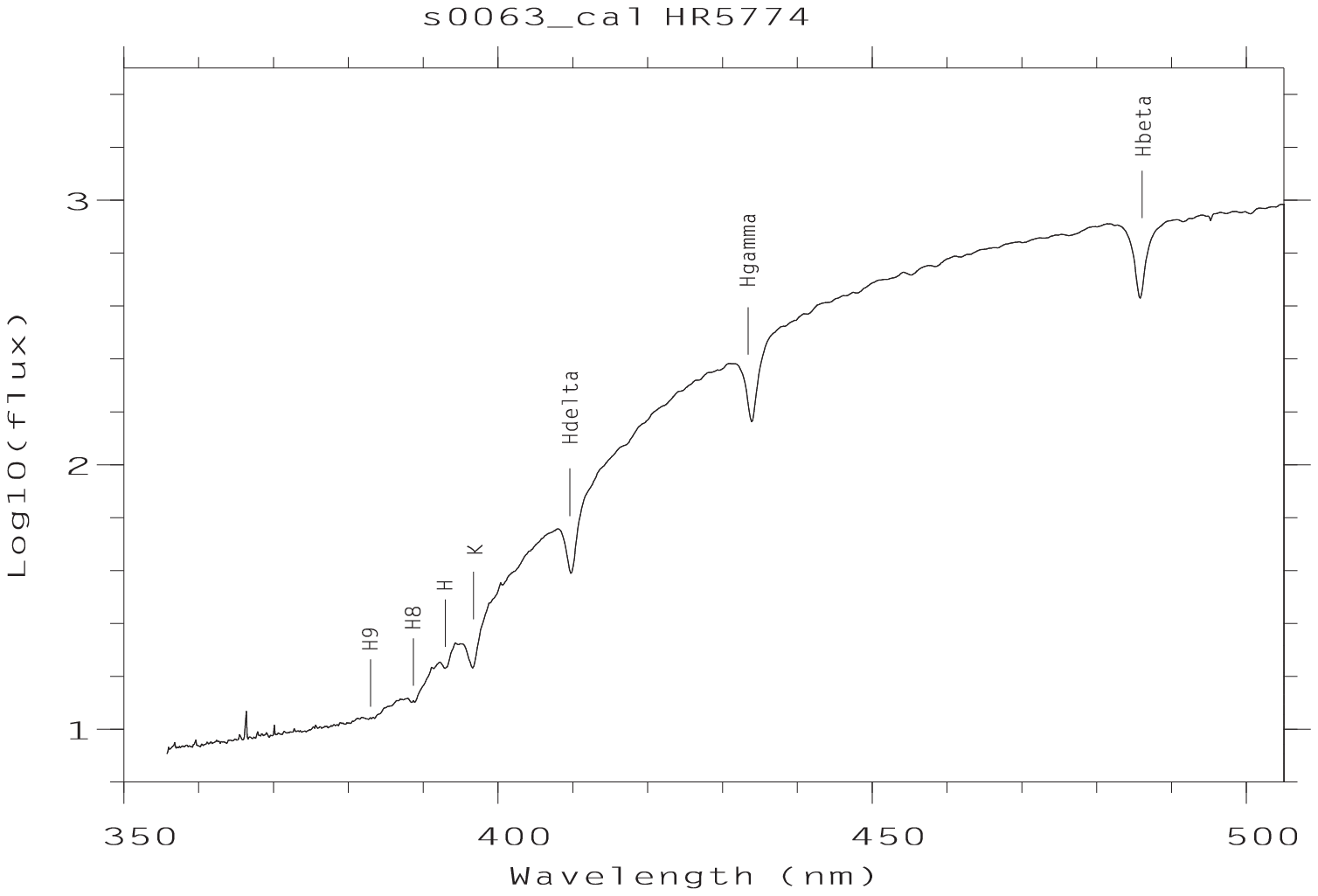}}
\smallskip
\centerline{Figure 7: First spectrum obtained with PISCO (HD 5774).}
%%%%%%%%%%%%%%%%%%%%%%%%%%%%%%%%%%%%%%%%%%%%%%%%%%%%%%%%%%%

\bigskip
\noindent{\large\sl 3.5. Wavefront analyzer}

\medskip
The atmospheric wavefront can be sensed
with the Shack-Hartman method (Roggeman {\it et al.}, 1997),
by putting a microlens array into the pupil plane (P1) and 
a specific imaging lens in the GR wheel, (cf. Fig.~\figopticdiag).
Each microlens has a diameter of 0.7~mm which
corresponds to 20~cm on the pupil at the TBL
and 10~cm at the 3.6~m CFH or ESO telescopes.

The SCIDAR (SCintillation Detection And Ranging) technique 
(Vernin \& Roddier, 1973) 
which consists in analysing the images of the pupil brightness
lit by two stars
to measure the wind speed and the altitude
of the turbulence layers,
could also be applied with PISCO.
In that case, fast detectors
operating at frequencies larger than 200~Hz
(such as the PAPA or the RANICON cameras, cf. \S 4) 
are needed to "freeze" the turbulence. 

\bigskip
\noindent{\large\bf 4. The detectors}

\penalty 10000
\medskip
PISCO has been used with a wide range of detectors.
Actually, the qualities of the detector mainly condition 
the performances of the image restoration process. A good knowledge of
the limitations of the detector is essential to elaborate a strategy 
of observation and obtain valid measurements.
Here we describe the detectors that we have already tested on PISCO.
For a wider information about the detectors used in the field of
optical speckle interferometry, see for instance a review in Cuby (1988), 
Richard {\it et al.}, (1990) or Cuby {\it et al.}, (1990). 

\bigskip
\noindent{\large\sl 4.1. The CP40-INSU detector}

\penalty 10000
\medskip
The first CP40 detector was designed 
at CERGA for interferometry observations by
A.~Blazit (Blazit, 1976, 1987). It is a two-stage intensified CCD camera
followed by a photon analyzer which computes
the coordinates of the photo-events (Foy, 1988b).
The field is covered by a mosaic of 4 CCD's, 288$\times$384 pixels each.
We actually used the duplication of this prototype, 
financed by INSU to make it available to the French astronomical 
community and in particular to the instruments of the TBL. 

The exposure time is set to 20~msec,
which may be too large for speckle applications
when the coherence time is smaller than this value.
To circumvent
this difficulty a rotating shutter was implemented
which reduces the exposure time to 5 or 10~msec. This shutter 
interrupts the light beam in the speckle camera
with a rotating opaque sector
synchronized with the frame signal of the CP40
(phase-locked motor). 

Because of a dead zone between the 4 image quadrants of the CP40, we 
decided to use only one quadrant and made a special
``off-axis'' mechanical interface 
to align the center of the selected quadrant
with the optical axis of PISCO. 

The geometrical distortion caused by the two-stage
amplifier is rather large, of the order of 20\% in the edges
(Thiebault, 1994).

Another problem is a strongly non-uniform sensitivity of the
photo-cathode within a single quadrant (down to nearly zero in one edge), 
which can hardly be corrected by a flat field map and 
causes a big non-uniformity of the signal to noise ratio 
within the elementary frames. The photometry of the image 
restoration process is also badly affected for intrinsically 
big objects that spread on the whole image.  

\bigskip
{\sl The ``photon-counting hole'' problem with the CP40.} 

\penalty 10000
\medskip
The electronic device which computes the coordinates
of the photo-events produces an artifact which
affects the photometry of the images.

When two photo-events are very close in the image, they merge into a
single spot. The photon centering device is unable to identify
it properly and discards such an event. This causes a
depletion of high spatial frequencies in the power spectrum.  
A ``hole'' can be seen in the center of the mean auto-correlation,
which becomes larger when the photon flux increases.
This problem also affects the photometry since many photons 
are not recorded in the high intensity regions of the image. 

To reduce this effect during our observations, the photon flux had to be 
limited to around 10,000 photons/sec and a high magnification was used to
over-sample by a factor of~3.

\bigskip
\noindent{\large\sl 4.2. The Ranicon} 

\penalty 10000
\medskip
The Ranicon (``Resistive Anode camera'', described 
in Clamping and Paresce (1989)) has been built by the Space
Telescope Science Institute (Baltimore). The model we used  
was lent by the Observatoire de la C\^ote d'Azur (OCA) 
for some observing runs between 1993 and 1996.

This detector has a S20 photo-cathode and a saturated mode single
microchannel amplifier (Gen II). The position analysis of the detected
photo-events is made with a resistive anode.
Each cloud of amplified electrons, resulting from the impact
of a photon on the photo-cathode, produces a charge drift
towards the four electrodes which surround the resistive anode. The location
of the impact is deduced from the voltage variations et the electrodes and
can be measured accurately to about~10~kHz.

Compared to the CP40, the quantum efficiency of the Ranicon is smaller by
a factor of $\sim$3. This is due to the lower efficiency of the GEN II
compared to the GEN I intensifiers.

Although the micro-channel amplification does not introduce
geometric distortion, we have noticed a small distortion
with the X and Y axes which are not perfectly perpendicular. A small
variation of the geometric scale was also noticed 
and calibration was needed during the night. 

Another unexpected defect was the presence of a small ``hole''
at the center of the autocorrelation function, similar
in some way to that of the CP40 (cf, \S 4.2), but with a
smaller amplitude. This is caused by the depletion of electrons
of a micro-channel after a photon-detection:
a delay of a few tenths of milliseconds is needed to recover its
full charge and efficiency. To reduce this defect,
the photon flux had to be lowered to about 8000 photons/sec.

\bigskip
\noindent{\large\sl 4.3. Other detectors} 

\penalty 10000
\medskip
Two other detectors have been used with PISCO:
the ICCD (Intensified Charge Coupled Device)
belonging to C.~Aime and E.~Aristidi's 
from Nice University, 
and P.~Nisenson's PAPA camera from Harvard Smithsonian
Center for Astrophysics (CfA):

\noindent
-- The ICCD has a single stage intensifier. It cannot
operate in true photon counting mode and is thus limited to objects brighter
than V$_{lim} \sim$10. This detector has no significant geometric 
distortion nor
non-linearity problems which would affect the photometry measurements. The
exposure time can be set between 64~$\mu$sec and 16~msec and the gain of the
micro-channel amplifier can be tuned, thus allowing a wide range of
input luminosities. The output is an analog video signal,
recorded on SVHS video cassettes. 
Bispectral image restoration with this detector has been very promising, 
and the first attempts lead to the restoration of a triple star 
(Aristidi {\it et al.}, 1997).

\noindent
-- The principle of the PAPA camera was described
in Papaliolios and Mertz (1982) and Papaliolios {\it et al.}, (1985). 
It features a two-stage electrostatic amplifier, and
a fast (P46) phosphor. Amplified photon impacts are analyzed by a
set of binary masks which act as an optical computer to instantly
digitize the position of photons in the field.  
The version we used was new and not fully operational, with
a new binary mask setup and 
a refurbished image intensifier jointly made by
P.~Nisenson, D.~Gezari (CfA) and  L.~Koechlin (OMP).
The first observations
in June 1997 have shown that the quantum efficiency was very good,
slightly larger than that of the CP40. The maximum photon
rate per second was as high as 100,000 but a small ``hole'' at the center of the
autocorrelation function was also noticed. The geometric distortion
caused by the image intensifier was large and an overall scale
variation during the night imposed quasi permanent scale calibrations. 
To solve this problem, the image intensifier was changed
after these observations.

\bigskip
\noindent{\large\sl 4.4. Comparison of the detectors} 

\penalty 10000
\medskip
Here is a summary of the characteristics of 
the detectors used with PISCO.

\penalty 10000
\medskip
\begin{enumerate}
\item The CP40 has a good quantum efficiency but a non-uniform
sensitivity and a strong geometric distortion, with a fixed
integration time of 20~msec.
The ``photon-counting hole'' affects the photometry and limits the
photon flux to around 10000~ph/sec (limiting magnitude
at the TBL: V$_{lim}\sim$12). 
\item The Ranicon has a very low geometric distortion, but a poor quantum
efficiency and a limitation of the usable photon flux to around 8000~ph/sec
(V$_{lim}\sim$11). 
It generates a chronologically ordered list of photon coordinates.
\item The PAPA exhibits a flat-field pattern and geometric distortion.
The photon flux is limited to around 100,000~ph/sec,
and V$_{lim}\sim$12. 
It also a chronologically ordered list of photon coordinates.
\item The ICCD of Nice Univ. has a lower gain than the previous detectors,
no geometric distortion, virtually no limitation to the photon flux
for normal astrophysical use, and V$_{lim}\sim$10. The image rate
is 50~Hz, with an electronic shutter, able to reduce the integration time
to 0.06 msec
\end{enumerate}

\penalty 10000
\medskip\noindent
Hence the detectors should be chosen according to the observing program, 
since some defects may be incompatible with the observation requirements.
A good detector for high resolution imaging is still to be desired.
Some technical developments are under way in our team to contribute
to this problem. A prototype of a new photon-counting camera
that would allow a high photon rate and a direct numerisation of photon coordinates
 is beeing tested (DELTA camera, Koechlin \& Morel, 1998).

\bigskip
\noindent{\large\bf 5. Performances and limitations}

\penalty 10000
\medskip
\noindent{\large\sl 5.1. Physical limitations}

\penalty 10000
\medskip
The effects of seeing on speckle observations is a strong reduction
of the limiting magnitude for bad seeing conditions,
whereas the angular resolution attainable in the image restoration process
degrades more slowly from the theoretical limit of
$\lambda/D$, where $\lambda$ is the wavelength
and $D$ is the telescope diameter (Roddier, 1981).

As solar observations have demonstrated, 
the Pic  du Midi site sometimes features slow seeing variations and extended 
isoplanetic patch, which indicates that the~TBL is potentially well suited
to  speckle and adaptive optics observations at short wavelengths, 
despite its modest size. With
a diffraction limit at 0.06\arcsec$\,$ in V, the~TBL can provide high
quality data from which many  astrophysical programs could benefit.

Due to the necessary short exposure times 
photon noise is the most  severe limiting factor in speckle imaging 
(Dainty and Greenaway, 1979, Beletic and Goody, 1992).
The limiting magnitude depends on the atmospheric seeing, the spectral
bandpass, the angular resolution  to be achieved, and the quantum efficiency
of the detector (Dainty and Greenaway, 1979).
We reach \hbox{$V \sim$12} with the TBL and detectors
such as the CP40 and the PAPA with a bandpass of 70~nm. 
Without filters, the expected limiting magnitude would
increase to about \hbox{$V \sim$14}.
In that case, the wavelength range is determined by the product
of the sensitivity response of the detector with that of the photo-cathode, 
and the resulting bandwidth is a few hundreds of nanometers.
It was shown both experimentally (Hege {\it et al.}, 1981)
and with numerical simulations
(Ziad {\it et al.}, 1994) that such extreme observing 
conditions can be used for faint detection of object duplicity.

\bigskip
\noindent{\large\sl 5.2. Data processing and image restoration}

\penalty 10000
\medskip
Whereas data quality is of paramount importance 
and obviously limits the angular resolution that
can be ultimately obtained in the reconstructed images, 
the nature of the data reduction methods subsequently 
employed to extract the scientific information contained 
in these images plays a key role in ensuring the overall 
success of the scientific programs. 

We have written the software to process
data from the various detectors used (cf.~\S 4),
and the different observing modes: speckle imaging, 
aperture synthesis with pupil masks, speckle spectroscopy,
and coronagraphy. The analysis of the wavefront is not yet 
implemented.

For speckle imaging, a few observers have 
independently reached the conclusion that ``the bispectrum 
combined with a constrained iterative deconvolution of 
amplitudes produces the highest quality imagery'' 
(Beletic and Goody, 1992). Nevertheless, we have used various 
programs ranging from Knox-Tompson (1974) to full bispectrum
methods (Weigelt, 1977, Roddier, 1986, Lannes, 1989) 
(and even partial bispectrum methods, i.e. using
only a subset of all possible closure relations) and found
little differences on the restored phasor image
of double stars. The pre-processing of the original data (correction
of geometric distortion, of flat-field, 
and various calibrations) is for us
the crucial step in the whole image restoration process.

The ``Aperture Synthesis'' team at OMP has been mainly involved 
during the last few years in the theoretical aspects of aperture 
synthesis and related problems such as deconvolution, 
wavelets and multi-resolution methods, with applications
on single aperture interferometry and multi-aperture devices
(Lannes {\it et al.\/} 1987; Lannes 1988, 1989, 1991).
The approach to these 
problems is deterministic and based on a least-squares 
scheme that allows error analysis, hence a good understanding 
of the stability of the image restoration process. 
Note that the deconvolution method (Lannes {\it et al.\/} 1987)
respects the photometry of the target, 
which is necessary for many applications, such as the determination of
color indices of binary stars for example (cf~\S 6).

\bigskip
\noindent{\large\bf 6. Astrophysical programs}

\penalty 10000
\medskip
In this section, we describe are some of the scientific 
subjects which are being studied with PISCO at Pic du Midi.
Other programs which aimed at imaging complex objects
(asteroids and stellar enveloppes)
have been impossible to do because of bad weather conditions.
    
\penalty 10000
\medskip\noindent
{\sl -- Orbits of binary stars}

\penalty 10000
\medskip
The study of binaries is a well suited program for speckle cameras 
as the bibliography of the last two decades can easily show
(CHARA project, McAlister and Hartkopf, 1984, 1988, 
Hartkopf {\it et al.}, 1996).

A long term program aims at measuring the position of close binaries 
to determine the
orbits and derive the masses of the components using the parallaxes 
measured by Hipparcos (Carbillet {\it et al.}, 1996, 
Aristidi {\it et al.}, 1997, Aristidi {\it et al.}, 1998).
New orbital elements have already been recalculated for
8~double stars from these observations
(Aristidi {\it et al.}, 1998).

We noticed that PISCO was very efficient for
binary study, even when the atmospheric conditions
were poor and did not allow any other imaging program. 
Hence binary measurements have been used as a backup program of
all our high angular resolution observations.

\bigskip\noindent
{\sl -- Stellar classification of components of binary stars}

\penalty 10000
\medskip
When images have been restored in B, V, R, with a good photometry
(cf. \S 5.2), color indices 
can be measured which then allow stellar classification of each of
the two stars. This may reveal essential for the stars for which 
a derivation of mass has been made. Acurate orbit determination
(and hence masses) are easier to perform for
short-period binaries, which are generally very close and for
which only global color indices or spectra are available.
The individual stellar classification is then poorly known.
The paradox is then that acurate masses are affected to
stars with big uncertainties in the stellar classification,
or less acurate masses to well identified stars, in the case
of binaries with a big angular separation. Hence we see that
color indices, or even spectra, of individual stars are
crucial for stellar studies.

A study of composite spectrum stars
(coll. J.-M.~Carquillat and N.~Ginestet, OMP)
associate the imaging and spectroscopic modes of PISCO.
Some stars exhibit the signature of a
composite spectrum which could be interpreted
as the sum of (at least) two spectra of different type.
(Ginestet {\it et al.}, 1994)
The aim of this program is to detect the possible presence of
a companion and then to identify the spectral type  
of both components either with color indices or
(when possible) with a spectrum with high angular
resolution which would separate the spectra
of the individual stars.

\bigskip\noindent
{\sl -- Search for binarity and statistical studies}

\penalty 10000
\medskip
The influence of the presence of a companion for star formation
(accretion of a disk) and stellar evolution (stellar winds, f.i.)
is not yet fully understood. Hence some surveys have been
undergone to determine the frequency of binarity among pre-main sequence
or post-AGB stars to constrain the theoretical models
with these statistical results. 
Once the binarity has been established, the next step is to identify 
the nature of each companion, either with 
its photometry or directly by spectroscopy.

A statistical study of pre-main sequence stars
has been started in 1996 with
complementary high angular observations made at ESO and CFHT
with Adaptive Optics (AO) in the infra-red
(Bouvier {\it et al.}, 1996).

Another program directed by E.~Aristidi and B.~Lopez 
(Nice Univ., France)
aims at searching for binaries among Mira-type stars
(for which binarity has been suspected by Hipparcos), 
and studying the interaction between the envelope of the Mira
and the atmosphere of the companion (Lopez {\it et al.}, 1998).

\bigskip
\noindent{\large\bf 7. Conclusion}

\penalty 10000
\medskip
The speckle camera of Observatoire Midi-Pyr\'en\'ees
has been tested in all its operating modes
and is now qualified for routine scientific exploitation.
Its versatility with multi-mode observational 
possibilities makes it particularly well suited to testing the new
methods of image restoration and aperture synthesis.
The experience gained with pupil masks may
have direct applications for reducing data from optical
interferometric arrays.

The good performances of speckle methods for
binary star observations have lead to consequent 
orbit measurements during the last twenty years all around the
world
and PISCO has started to bring its contribution to this
effort (Carbillet {\it et al.}, 1996, 
Aristidi {\it et al.}, 1997, Aristidi {\it et al.}, 1998).
This high efficiency makes speckle observations a ``privileged'' tool
for binary studies. A new series of speckle programs
have been impulsed by the discovery by Hipparcos 
of thousands of binary candidates (confirmation of binarity, orbits, 
variability of companions, etc).
New space projects such as the space interferometers
dedicated to parallax measurements (ESA GAIA) will
also need follow-up based-ground observational programs
in the future for which PISCO and speckle techniques in general may
significantly contribute.

\bigskip
\noindent
{\large\bf Acknowledgements:}

\smallskip
We are indebted to A.~Blazit, D.~Mourard, E.~Aristidi, D.~Gezari
and P.~Nisenson for lending us their detectors, and to
A.~Lannes, M.~Festou, J.-M.~Carquillat, N.~Ginestet
and M.~Scardia
for the fruitful collaboration for the scientific exploitation
of PISCO.

We thank the Observatoire Midi-Pyr\'en\'ees technical staff,
and especially the workshop of Toulouse, Bagn\`eres de 
Bigorre and Pic du Midi and the night assistants and operators 
of the TBL, for their participation to this project. 
We acknowledge the assistance of J.~Cadaugade and S.~Chastanet for the 
preparation of the photographs. 

This instrument was financed by a grant from the 
{\sl Institut National des 
Sciences de l'Univers} of the {\sl Centre National de la Recherche 
Scientifique (CNRS)} to the~TBL, with additional support from 
the {\sl Unit\'es de Recherche Associ\'ees n$^{\circ}$1281}
and {\sl n$^{\circ}$285} (now {\sl Unit\'e Mixte de Recherche 
n$^{\circ}$5572}) of CNRS.
 
\bigskip
\centerline{\large\bf Bibliography}
 
\medskip
{\parindent0pt

\bibitem{} Andr\'e,~C., Festou,~M.C., Koechlin,~L., Lannes,~A., Perez,~J.-P., 
Prieur,~J.-L., Roques,~S., 1994,
\fullref{High resolution imaging of comets and asteroids using 
bispectral techniques,} 
Planet. Space Sci., 42, 747--758.

\bibitem{} Anterrieu,~E., 1992,
\fullref{Synth\`ese d'ouverture multi-pupillaire. 
Algorithmique de reconstruction d'image,}
Thesis, Univ. Paul Sabatier, Toulouse, France.

\bibitem{}  Aristidi, E., Carbillet, M., Prieur, J.-L, Lopez, B., Bresson, Y., 
1997, 
\fullref{ICCD speckle observations of binary stars: measurements 
during 1994--1995} 
A\&A Suppl., 126, 555.

\bibitem{}  Aristidi, E., Prieur, J.-L., Scardia, M., Koechlin, L.,
Avila, R., Lopez, B., Rabbia, Y., Carbillet, M., Nisenson, P., Gezari, D.,
1998, 
\fullref{Speckle observations of double and multiple stars at Pic du Midi: 
measurements during 1995 and 1997, and orbits.} 
A\&A Suppl., accepted. 

\bibitem{} Beckers, J.M., Hege, E.K., Murphy, H.P., 1983,
Proc. Soc. Photo-Opt. Instrum. Eng., 445, 462.
%% Steward speckle camera, described in Kuwamura et al, 1993

\bibitem{} Beckers, J.M., Merckle, F., 1991, 
Proc. of the ESO conference on:
High Resolution Imaging by interferometry II,
Garching, Germany.

\bibitem{} Beletic, J.W., Goody, R.M., 1992,
\fullref{Limits to the recovery of planetary images by speckle imaging,}
Applied Optics, 31, 6909--6921.

\bibitem{} Blazit, A., 1976, 
Thesis, University of Paris VII, France.

\bibitem{} Blazit, A., 1987, 
PhD Thesis, University of Nice, France.

\bibitem{} Blazit, A., Bonneau, D., Koechlin, L., Labeyrie, A., 1977,
\fullref{The digital speckle interferometer: preliminary results on 
59 stars and 3C273,}
Astroph. Journ., 214, L79--L84.

\bibitem{} Bouvier,~J., Corporon,~P., Prieur,~J.-L., Rigaut,~F, Tessier,~E.,
Brandner,~W., 1996,
\fullref{A diffraction-limited imaging survey of binaries in 
intermediary-mass pre-main sequence stars}
in: Planetary Formation in the Binary Environment, Stony Brook,
16--18~Juin 1996.

\bibitem{} Breckinridge, J.B., McAlister, H.A., Robinson, W.G., 1979,
\fullref{Kitt Peak speckle camera}
Appl. Optics, 18, 7, 1034--1041. 

\bibitem{} Carbillet, M., Lopez, B., Aristidi, E., Bresson, 
Y., Aime, C., Ricort, G., Prieur, J.-L., Koechlin, L., 
Helmer, G., Lef\`evre, J., Cruzalebes, P., 1996, 
\fullref{Discovery of a new bright close double star} 
A\&A, 314, 112.

% Ranicon camera
\bibitem{} Clamping, F., Paresce, F., 1989, 
A\&A, 225, 578--584.

\bibitem{} Cuby, J.-G., 1988, Ann. Phys., 13.

\bibitem{} Cuby, J.-G., et al., 1990, SPIE 1235, p. 294.

\bibitem{} Dainty, J.C., 1984,
\fullref{Stellar speckle interferometry,}
in Topics in Applied Physics, {\it Laser speckle and related 
phenomena} Vol. 9, (Springer-Verlag) 258--320.

\bibitem{} Dainty, J.C. and Greenaway, A.H., 1979,
\fullref{Estimation of spatial power spectra in speckle interferometry,}
JOSA, Vol. 69, N$^o$5, 786. 

\bibitem{} Ginestet N., Carquillat J.M., Jaschek M., Jaschek C., 1994, 
\fullref{Spectral classifications in the near infrared 
of stars with composite spectra. I.~The study of MK standards.}
A\&A Suppl., 108, 359--375.

\bibitem{} Golay, M.J.E., 1971,
\fullref{Points arrays having compact, nonredundant autocorrelations,}
JOSA, 61, 272--273.

\bibitem{}  Foy R., 1988a,
Speckle Imaging Review.
In: Instrumentation for ground-based optical astronomy, present and future,
Robinson L.B. (eds), Springer-Verlag (New-York), p345

\bibitem{}  Foy R., 1988b,
The Photon-Counting Camera CP40.
In: Instrumentation for ground-based optical astronomy, present and future,
Robinson L.B. (eds), Springer-Verlag (New-York), p345

\bibitem{} Hege, H.K., Hubbard, E.N., Strittmatter, P.A., Worden, S.P.,
1981, Ap.J., 248, L1-L3.
%%%% Detection of duplicity of QSO PG 1115+08 in white light

\bibitem{} Hubbard et al.   A.J., 84, 1437.
%%%% SO/AFGL speckle camera

\bibitem{} Knox, K. T., Thomson, B. J., 1974,
\fullref{Recovery of images from atmospherically degraded 
short exposure photographs,}
Astroph. J., 193, L45--L48.

\bibitem{} Koechlin, L., \& Morel, S., 1998, A\&A, accepted.

\bibitem{} Kuwamura, S., Baba, N., Miura, N., Hege, E.K., 1993
\fullref{Stellar spectra reconstruction from speckle spectroscopic data,}
Astron.~J., 105, 665--671.

\bibitem{} Labeyrie, A., 1970,
\fullref{Attainment of diffraction limited resolution in large 
telescopes by Fourier analysing speckle patterns in star images,}
A\&A, 6, 85--87.

\bibitem{} Lannes, A., 1988,
\fullref{On a new class of iterative algorithms for phase-closure 
imagingand bispectral analysis,}
Proc. of the NOAO-ESO conference on:
High Resolution Imaging by interferometry, Garching, 169--180.

\bibitem{} Lannes, A., 1989,
\fullref{Back projection mechanisms in phase-closure imaging.
Bispectral analysis of the phase-restoration process,}
Exp. Astron., 1, 47--76.

\bibitem{} Lannes, A., 1991,
\fullref{Phase and amplitude calibration in aperture synthesis. 
Algebraic structures,}
Inverse Problems, 7, 261--298.

\bibitem{} Lannes, A., Roques, S., Cazanove, M.J., 1987a,
\fullref{Stabilized reconstruction in signal and image processing; 
Part~I: Partial deconvolution and spectral extrapolation with limited field,}
J. Mod. Optics, 34, 161--226.

\bibitem{} Lannes, A., Roques, S., Cazanove, M.J., 1987b, 
\fullref{Stabilized reconstruction in signal and image processing; 
Part~II: Iterative reconstruction with and without constraint. 
Interactive implementation,}
J. Mod. Optics, 34, 321--370.

\bibitem{} Lopez, B., Aristidi, E., Prieur, J.-L., {\it  et al.}, 1998,
in preparation.
% Mira 1995....

\bibitem{} McAlister H.A., Hartkopf W.I., 1984, Catalog of Interferometric 
Measurements of Binary Stars, CHARA contr. n$^\circ$1, 
(Georgia State Univ., Atlanta).

\bibitem{} McAlister H.A., Hartkopf W.I., 1988, 
Second Catalog of Interferometric 
Measurements of Binary Stars, CHARA contr. n$^\circ$2,
(Georgia State Univ., Atlanta).

\bibitem{} Hartkopf, W.I., Mason, B.D., McAlister H.A., Turner, N.H.,
Barry, D.J., Franz, O.G., Prieto, C.M., 1996
\fullref{ICCD Speckle Observations of Binary Stars. XIII. Measurements
during 1989---1994 from the Cerro Tololo 4-m Telescope}
A.J., 111, 936.

\bibitem{} Owens,~J.C., 1967, 
\fullref{Optical refractive index of air: dependence on 
pressure, temperature and composition,} 
Applied Optics, Vol.~6, N$^\circ$1., 51--59.

% PAPA camera
\bibitem{} Papaliolios, C., Mertz, L., 1982,
Proc. SPIE, 331, 360. 

\bibitem{} Papaliolios, C., Nisenson, P., Ebstein, S., 1985, 
\fullref{Speckle imaging with the PAPA detector}
Appl. Optics, 24, 287--292.

\bibitem{} Prieur, J.-L., Lannes, A., Cullum, M., 1991,
\fullref{Image reconstruction using bispectral techniques,}
Proc. of the ESO conference on: 
High Resolution Imaging by interferometry II,
(Ed. J.M.~Beckers and F.~Merckle) 
Garching, Germany, 353--357.

\bibitem{} Prieur,~J.-L., Festou,~M.C, Koechlin,~L., Andr\'e,~C., 
1994,
\fullref{Le tavelographe \`a masques 
pupillaires de l'OMP,}
Coll. {\sl National de Plan\'etologie de l'INSU},
Toulouse, 13--16~Juin 1994,
S16-22.

\bibitem{} Richard, J.-C., et al., 1990, SPIE 1235, p. 294.

\bibitem{} Roddier, F., 1981,
\fullref{The effects of atmospheric turbulence in optical astronomy,}
Ed. E. Wolf, Progress in optics XIX North-Holland, Vol.~2, 
p~283.

\bibitem{} Roddier, F., 1986,
\fullref{Triple correlation as a phase closure technique,}
Opt. Comm., 60, 145--148.

% Phase mask in the image plane with a delay of PI:
\bibitem{} Roddier, F., \& Roddier, C., 1997,
PASP, 109, 815--820.

\bibitem{} Roques, S., 1987,
\fullref{Probl\`emes inverses en traitement d'image, r\'egularisation
et r\'esolution en imagerie bidimensionnelle,}
Th\`ese d'Etat. Univ. Paul Sabatier, Toulouse, France.

%Shack Hartman method among many other things such as adaptive optics,
\bibitem{} 
Roggeman, M.C., Welsh, B.M., Fugate, R.Q., 1997,
\fullref{Improving the resolution of ground-based telescopes}
Rev. of Modern Physics, Vol 69, 2, 437--505.

\bibitem{} Simon, G.W., 1966,
\fullref{A practical solution of the atmospheric dispersion problem,}
Astron.~J., {\bf 71}, 190.

\bibitem{} Strittmatter, P.A., 1980,
\fullref{The Steward speckle interferometry and speckle holography 
program,}
in: Applications of speckle phenomena,  SPIE, 243, 103-111.

\bibitem{} Thiebaut, E., 1994,
\fullref{Imagerie astrophysique \`a la limite de diffraction 
des grands t\'elescopes.  Application \`a l'observation des objets froids.}
Thesis, Univ. Paris VII, France.

\bibitem{} Vernin, J., Roddier, F., 1973, JOSA, 63, 270.

\bibitem{} Weigelt, G., 1977,
\fullref{Modified astronomical speckle interferometry, speckle 
masking,}
Opt. Comm., 21, 55--59.

\bibitem{} Weigelt,~G, Grieger,~F., Hofmann,~K.-H., Pausenberger,~R., 
1991, Colloque ESO.
\fullref{Slit Speckle Spectroscopy}
{\it High Resolution Imaging by interferometry II}, p.471.

\bibitem{} Wallner, E.P., Wetherell, W.B., 1990,
%\fullref%%%%%
{\sl Broad spectral bandpass atmospheric dispersion correctors,}
Rapport technique Itek.

\bibitem{} Ziad, A., Borgnino, J., Agabi, A., Martin, F., 1994,
\fullref%%%%%
{Optimized spectral bandwidth in high angular resolution imaging} 
Exp. Astron., 5, 247-268.

}%%%%% end of block with parindent=0pt.

\end{document}